\documentclass[aps,prl,reprint,floatfix]{revtex4-2}

\usepackage{amsmath}
\usepackage{amssymb}
\usepackage{mathtools}
\usepackage[bb=dsserif]{mathalpha}
\usepackage{bm}
\usepackage[per-mode=symbol, uncertainty-mode=compact-marker]{siunitx}
\DeclareSIUnit\bar{bar}
\DeclareSIUnit\amu{amu}
\DeclareSIUnit\eVpercSquared{\text{eV}/\text{\ensuremath{c^2}}}
\usepackage{isotope}
\usepackage{graphicx}
\usepackage[nomain,acronym,postdot,nogroupskip]{glossaries-extra}
\usepackage{longtable}
\usepackage{multirow}
\usepackage{booktabs}
\usepackage{xcolor}
\usepackage[hidelinks]{hyperref}

\newcommand{\term}[3]{\ensuremath{{}^{#1}\mathrm{#2}_{#3}}}
\renewcommand{\to}{\ensuremath{{\rightarrow}}}

\newabbreviation{MOT}{MOT}{magneto-optical trap}
\newabbreviation{SM}{SM}{standard model}
\newabbreviation{BSM}{BSM}{beyond standard model}
\newabbreviation{IS}{IS}{isotope shift}
\newabbreviation{mIS}{mIS}{mass-normalized isotope shift}
\newabbreviation{ISS}{ISS}{isotope shift spectroscopy}
\newabbreviation{KP}{KP}{King plot}
\newabbreviation{GKP}{GKP}{generalized King plot}
\newabbreviation{QCD}{QCD}{quantum chromodynamics}
\newabbreviation{CKM}{CKM}{Cabibbo-Kobayashi-Maskawa}
\newabbreviation{CP}{CP}{charge and parity symmetry}
\newabbreviation{CPT}{CPT}{charge, parity and time reversal symmetry}
\newabbreviation{LHC}{LHC}{Large Hadron Collider}
\newabbreviation{HV}{HV}{high vacuum}
\newabbreviation{UHV}{UHV}{ultra-high vacuum}
\newabbreviation{XHV}{XHV}{extreme high vacuum}
\newabbreviation{UV}{UV}{ultraviolet}
\newabbreviation{IR}{IR}{infrared}
\newabbreviation{PDH}{PDH}{Pound-Drever-Hall}
\newabbreviation{AOM}{AOM}{acousto-optic modulator}
\newabbreviation{EOM}{EOM}{electro-optic modulator}
\newabbreviation{PBS}{PBS}{polarization beam splitter}
\newabbreviation{RF}{RF}{radio frequency}
\newabbreviation{PID}{PID}{proportional-integral-differential}
\newabbreviation{ULE}{ULE}{ultra-low expansion}
\newabbreviation{ECDL}{ECDL}{external-cavity diode laser}
\newabbreviation{CCD}{CCD}{charge-coupled device}
\newabbreviation{PD}{PD}{photodiode}
\newabbreviation{FS}{FS}{field shift}
\newabbreviation{QFS}{QFS}{quadratic field shift}
\newabbreviation{MS}{MS}{mass shift}
\newabbreviation{FSR}{FSR}{free spectral range}
\newabbreviation{NMS}{NMS}{normal mass shift}
\newabbreviation{SMS}{SMS}{specific mass shift}
\newabbreviation{MCDF}{MCDF}{multiconfiguration Dirac-Fock}
\newabbreviation{MCDHF}{MCDHF}{multiconfiguration Dirac-Hartree-Fock}
\newabbreviation{ND}{ND}{nuclear deformation}
\newabbreviation{NP}{NP}{nuclear polarization}
\newabbreviation{QED}{QED}{quantum electrodynamics}
\newabbreviation{Hg}{Hg}{mercury}
\newabbreviation{Ca}{Ca}{calcium}
\newabbreviation{Yb}{Yb}{ytterbium}

\begin{document}


\title{Probing Nuclear Interactions Through Isotope Shift Spectroscopy of Mercury} 

\author{Thorsten Groh}
\email{Contact author: groh@physik.uni-bonn.de}
\affiliation{Physikalisches Institut, Universität Bonn, Nussallee 12, 53115 Bonn, Germany}

\author{Felix Affeld}
\affiliation{Physikalisches Institut, Universität Bonn, Nussallee 12, 53115 Bonn, Germany}

\author{Simon Stellmer}
\affiliation{Physikalisches Institut, Universität Bonn, Nussallee 12, 53115 Bonn, Germany}

\date{\today}

\begin{abstract}
We present precision isotope shift spectroscopy of the $\mathrm{6s^{2}}\, \term{1}{S}{0}{\to\,}\mathrm{6s\, 6p}\, \term{3}{P}{1}$ intercombination line and the $\mathrm{6s\,6p}\, \term{3}{P}{1}{\to\,}\mathrm{6s\,6d}\, \term{3}{D}{J}$ ($J=1,2$) transitions in neutral \glsxtrlong{Hg}, performed on the five naturally abundant even isotopes, including the low-abundant isotope \isotope[196]{Hg}.
Using laser-cooled atoms in a \glsxtrlong{MOT}, we achieve uncertainties down to \SI{20}{\kilo\hertz}, resolving the \glsxtrlongpl{IS} to a fractional uncertainty of $\sim\num{2e-6}$.
A \glsxtrlong{KP} analysis comparing our $\term{1}{S}{0}{\to}\term{3}{P}{1}$ data to previous results on the $\mathrm{6s\,6p}\,\term{3}{P}{2}{\to\,}\mathrm{6s\,7s}\,\term{3}{S}{1}$ line reveals a nonlinearity with $4.9\,\sigma$ significance.
Our \glsxtrlong{GKP} nonlinear decomposition analysis discusses potential contributions from quadratic ($\propto \delta\langle r^{2}\rangle^{2}$) and higher order field shifts ($\propto \delta\langle r^{4}\rangle$) also induced by \glsxtrlongpl{ND}.
These measurements yield new insights into the structure of the \glsxtrshort{Hg} nucleus and provide benchmarks for nuclear-structure models.
They further establish \glsxtrlong{Hg} as a potential platform to search for hypothetical Yukawa-type boson-mediated forces coupling electrons to neutrons.
\end{abstract}

\maketitle

High-precision, low-energy experiments now reach extraordinary accuracy, with optical clocks operating at the \num{e-18} level~\cite{Brewer2019, Aeppli2024, McGrew2018}, hydrogen spectroscopy testing theory at \num{e-15}~\cite{Parthey2011, Matveev2013}, and the electron $g$-factor known at the level of \num{e-13}~\cite{Fan2023, Hanneke2008}, enabling stringent tests of fundamental symmetries, dark matter searches, and searches for variations of fundamental constants~\cite{Safronova2018, Roussy2021, Derevianko2014}.

Recently, \gls{ISS}\glsunset{IS} has emerged as a key probe at the interface of atomic, nuclear, and particle physics~\cite{Counts2020, Figueroa2022, Ono2022, Hur2022, Door2025, Ishiyama2025, Gebert2015, Roeser2024, Wilzewski2025, Hofsaess2023, Roeser2024, Rehbehn2023}.
In particular, \gls{ISS} provides sensitivity to \gls{BSM} parity-conserving electron-neutron interactions, mediated by hypothetical light bosons~\cite{Delaunay2017, Berengut2018}. 
This search is also motivated by the reported \SI{17}{MeV} \isotope[8]{Be} anomaly, interpreted as a potential vector boson~\cite{Krasznahorkay2016, Feng2016}.  
Such interactions are inaccessible to collider experiments, but could manifest as nonlinearities in \gls{KP} analyses of \glspl{IS}~\cite{Berengut2025}. 
Recent high-resolution \gls{IS} measurements in \gls{Ca} and \gls{Yb} provide strong bounds on a Yukawa-type electron-neutron  interaction across boson masses from \SIrange[range-units = single]{10}{e7}{\eVpercSquared}~\cite{Solaro2020, Chang2024, Wilzewski2025, Counts2020, Figueroa2022, Ono2022, Hur2022, Door2025, Ishiyama2025} and exclude wide regions of parameter space associated with extended Higgs sectors, alternative mass-generation mechanisms, and light mediators~\cite{Delaunay2017}; with bounds now comparable to astrophysical and other constraints~\cite{Bordag2009, Grifols1989, Fan2023, Raffelt2012, Nesvizhevsky2008, Delaunay2017a}.

\textit{Isotope shift---}The \gls{IS}, $\delta\nu_i^{A-A'} \equiv \nu_i^{A} - \nu_i^{A'}\hspace{-0.17em}$, is the change in the frequency $\nu_i$ of an electronic transition~$i$ between isotopes of mass numbers $A$ and $A'$.
To leading order perturbation theory, we can write
\begin{equation}
  \delta\nu_i^{A-A'}\!\! = F_i\, \delta\!\left\langle r^2 \right\rangle^{A-A'}\!\!\! + K_i\,  \delta\mu^{A-A'}\! + \sum_{\kappa} G_i^{(\kappa)}\, \delta\eta_{(\kappa)}^{A-A'}\!\!,
  \label{eq:isotope_shift}
\end{equation}
where $F_i$, $K_i$ and $G_i^{(\kappa)}$ denote transition dependent electronic and the $\delta x^{A-A'} \equiv x^A - x^{A'}\hspace{-0.17em}$ nuclear factors \cite{Berengut2025}.

In heavy elements like \gls{Hg}, the \gls{IS} is dominated by the \gls{FS}, which is proportional to the difference in mean-square nuclear charge radii $\delta\langle r^2 \rangle^{A-A'} \equiv \langle r^2 \rangle^{A} - \langle r^2 \rangle^{A'}\hspace{-0.17em}$.
Originating from the overlap of the electronic wavefunction with the nuclear charge distribution, it is strongest for s-electrons: in \gls{Hg}, the $\mathrm{6s^2}$ ground state largely determines the \gls{IS} of the intercombination line~\cite{Zhang2019, Schelfhout2022}.
The \glsxtrfull{MS}, arising from the nuclear recoil, is proportional to the inverse nuclear mass difference, $\delta\mu^{A-A'} \equiv {1}/{m^A} - {1}/{m^{A'}}\hspace{-0.17em}$. 
It approximately scales as $1/({m^A})^2$, making it strongest in the hydrogen-deuterium \gls{IS} and dominates the IS in light systems, such as \gls{Ca}, while being strongly suppressed in \gls{Hg} \cite{King1984}.

The terms $\delta\eta_{(\kappa)}^{A-A'}\hspace{-0.17em}$ denote higher order (nuclear) corrections. 
For heavy elements, higher order nuclear moments, like the quadratic \glsxtrshort{FS}\glsunset{QFS} (QFS) $\propto \delta(\langle r^2 \rangle^2)^{A-A'}\hspace{-0.17em}$, are expected to be the next resolvable contributions \cite{Counts2020}. 
Recent \gls{ISS} measurements in \gls{Yb} ions on the sub-hertz level \cite{Counts2020, Figueroa2022, Ono2022, Hur2022, Door2025, Ishiyama2025} have revealed \gls{ND} contributions $\propto \delta \langle r^4 \rangle^{A-A'}\hspace{-0.17em}$ (= quartic \gls{FS} \cite{Door2025}) reflecting prolate deformation of the nuclear charge distribution limiting \gls{BSM} searches \cite{Berengut2025}.
By contrast, the near-spherical nuclei of \gls{Ca} suppress such corrections: early measurements confirmed linearity~\cite{Gebert2015, Roeser2024}, while combined \isotope{Ca^+} and \isotope{Ca^{14+}} spectroscopy later uncovered a significant \gls{KP} nonlinearity consistent with second-order \gls{MS} $\propto 1 / ({m^A})^2 - 1 / ({m^{A'}})^2$ and \gls{NP} effects~\cite{Wilzewski2025}, the latter being a subtle electron interaction induced deformation of the nuclear charge distribution.
For certain transitions, second-order \gls{FS} or \gls{MS} effects mix nearby levels, yielding large \gls{IS} corrections \cite{Berengut2020} that produced up to \SI{32}{\mega\hertz} shifts in samarium \cite{Griffith1979, Griffith1981, Palmer1982}.

Despite these \glsxtrshort{SM} contributions, \gls{ISS} can be used to probe for a new physics boson $\phi$ coupling neutrons to electrons via a Yukawa-like interaction, $V_\text{NP}(r){=-}\alpha_\text{NP}\, (A-Z)\, e^{-m_\phi\, r} / r$ \cite{Delaunay2017, Haber1979}.
This interaction introduces a term $\propto \alpha_\text{NP}\, X_i\, \gamma^{A-A'}\hspace{-0.17em}$ that only depends on the difference in the number of neutrons, $\gamma^{A-A'} = (A-A')$.
Precision \gls{ISS} can place bounds on the electron to neutron coupling strength $\alpha_\text{NP} = (-1)^s\,\gamma_e\,\gamma_n$ and the mediator mass $m_\phi$ \cite{Delaunay2017, Berengut2018}.

\textit{Mercury---}\Gls{Hg} ($Z=80$) provides a system capturing the strengths of \gls{Ca} and \gls{Yb}.
Its nuclei are close to spherical due to the nearby proton shell closure at $Z=82$ and neutron shell closure at $N=126$, reducing nuclear deformation-driven contributions \cite{Ebata2017, Sun2024}, while the large nuclear charge $Z$ enhances sensitivity to electron–neutron couplings.  
\gls{Hg} provides narrow optical clock transitions and five stable even-even isotopes (see~Tab.~\ref{tab:isotope_shifts}), enabling high-precision \gls{ISS} and \gls{KP} analysis. 
\Gls{Hg} \gls{IS} measurements range from early arc spectra and vapor cell studies~\cite{Blaise1957,Gerstenkorn1977,Kroll1982,Rayman1989,Sansonetti2010} to very recent comb-referenced Doppler-free saturation spectroscopy experiments at sub-\si{\mega\hertz} \cite{Witkowski2019} and down to \si{\kilo\hertz}-level \cite{Gravina2024, Gravina2025} precision. 
Apart from searches for new forces, \gls{Hg} spectroscopy benchmarks nuclear structure through precise charge radii, exhibiting features like the kink at $N=126$ and shape staggering in neutron-deficient isotopes~\cite{Angeli2013}.  
Atomic beam and collinear laser \gls{ISS} established charge radii and magnetic dipole and electric quadrupole moments of \isotope[181]{Hg} to \isotope[206]{Hg} \cite{Ulm1986}, recent studies at ISOLDE extended these measurements to isotopes as light as \isotope[177]{Hg}, revealing pronounced odd-even staggering consistent with shape coexistence~\cite{DayGoodacre2017, Sels2018, Sels2019, DayGoodacre2021, Marsh2022}.

\begin{figure}
 \includegraphics[width=85mm]{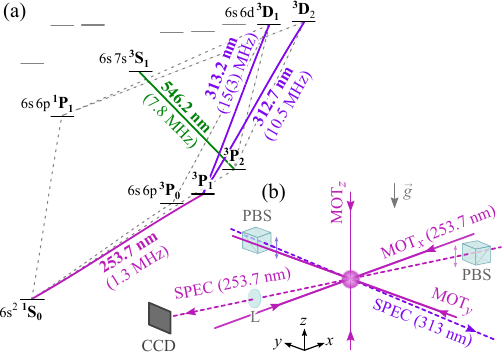}
 \caption{
    \label{fig:1}
    (a) Level scheme of \gls{Hg} relevant for \glsxtrlong{ISS} within this work (natural linewidths given in brackets).
    (b) Experimental setup for spectroscopy of the $\term{1}{S}{0}{\to}\term{3}{P}{1}$ intercombination line at \SI{253.7}{\nano\meter} and the $\term{3}{P}{1}{\to}\term{3}{D}{J}$ lines at \SI{313.2}{\nano\meter} ($J=1$) and \SI{312.7}{\nano\meter} ($J=2$).
    Atoms are cooled in a \gls{MOT} at \SI{253.7}{\nano\meter}.
    For resolving the $\term{1}{S}{0}{\to}\term{3}{P}{1}$ line, atoms are probed in free fall via absorption imaging at \SI{253.7}{\nano\meter}.
    The light is vertically polarized via a \glsxtrfull{PBS}, propagates in the $xy$-plane and is imaged on a \glsxtrshort{UV}-sensitive \glsxtrfull{CCD} sensor via a single objective lens (L).
    For resolving the $\term{3}{P}{1}{\to}\term{3}{D}{J}$ lines, vertically polarized light at \SI{313}{\nano\meter} depletes the \gls{MOT} atom number via optical pumping and subsequent decay to $\mathrm{{}^3P_{0,2}}$ and $\mathrm{{}^1P_1}$ (gray, dashed lines).
    Atom numbers within the \gls{MOT} are then determined by fluorescence imaging at \SI{253.7}{\nano\meter}.
  }
\end{figure}

\textit{Experiment---}We probe the \gls{IS} of neutral \gls{Hg} atoms on three distinct dipole-allowed lines in a \glsxtrfull{MOT} as illustrated in Fig.~\ref{fig:1}.
The setup is similar to the one described in Ref.~\cite{Lavigne2022}.
For the most abundant \isotope[202]{Hg}, we load up to $N=\num{1e7}$ atoms from a background gas and cool them to $T \approx \SI{150}{\micro\kelvin}$. 

The \gls{UV} spectroscopy light is derived via frequency conversion from an \gls{IR} \glsxtrlong{ECDL} at \SI{1015}{\nano\meter} ($\nu{\,\rightarrow\,}4\,\nu$) and \SI{1565}{\nano\meter} ($\nu{\,\rightarrow\,}5\,\nu$) respectively.
The quintupled \SI{313}{\nano\meter} laser system is similar to the setup described in Ref.~\cite{Bueki2021} and generates about \SI{1}{\milli\watt} of stable laser power in the \gls{UV}.
We lock the fundamental wavelength of the spectroscopy lasers to a length-stabilized \gls{ULE} cavity (finesse $\mathcal{F} = \num[group-separator = {\,}]{74100\pm 660}$ at $\SI{1015}{\nano\meter}$ and $\mathcal{F} = \num[group-separator = {\,}]{166460\pm 350}$ at $\SI{1565}{\nano\meter}$), which acts as the frequency reference for \gls{ISS}.
A fiber \glsxtrlong{EOM} shifts the laser frequency relative to the cavity resonances to address individual isotopes.
For $\term{1}{S}{0}{\to}\term{3}{P}{1}$, where the \gls{IS} spans about \SI{20}{\giga\hertz} in the \gls{UV} and exceeds the \gls{FSR} of the cavity ($\delta_\text{FSR} = \SI[group-separator = {\,}]{2992455.6 \pm 0.3}{\kilo\hertz}$), we lock the master laser to the nearest cavity resonance.
We characterize the ageing of the cavity spacer as a linear drift in time using repeated measurements of the  $\term{1}{S}{0}{\to}\term{3}{P}{1}$ transition.
A simultaneous fit, $\delta_\text{cav}^A(t) = n_\nu\, a_\text{cav}\, t + \delta_\text{cav}^{\star A}$, to all isotopes gives $a_\text{cav} = \SI{-0.200(4)}{\hertz\per\second}$ (at \SI{1015}{\nano\meter}) and the drift-corrected atomic resonance ($\nu_0^A$) to cavity reference line detuning $\delta_\text{cav}^{\star A} \equiv \nu_0^A - n_\nu\, \nu_{\text{cav}, 0}$, where $n_\nu=4$ ($n_\nu=5$) denotes the conversion factor from \gls{IR} to \gls{UV}.
\glsxtrlongpl{AOM} in the \gls{UV} scan the probe beam detuning $\Delta$ across the atomic linewidth and control the beam power.
All \glsxtrshort{RF} frequencies are phase locked to a \SI{10}{\mega\hertz} Rb microwave clock reference.

For spectroscopy of the $\term{1}{S}{0}{\to}\term{3}{P}{1}$ intercombination line at \SI{253.7}{\nano\meter}, where we also operate the \gls{MOT}, we probe the atoms in free fall via absorption imaging.
We switch off both the magnetic gradient field and the cooling light and spectroscopically address the atoms with a probe beam intensity of $s_0 = I / I_\mathrm{sat} \approx \num{0.02}$ for an exposure time of $t_\mathrm{exp} = \SI{150}{\micro\second}$.

For the $\term{3}{P}{1}{\to}\term{3}{D}{J}$ lines, we instead keep the \gls{MOT} cooling light on, which populates the excited $\mathrm{{}^3P_1}$ state.
Atom numbers are then determined via their fluorescence at \SI{253.7}{\nano\meter}.
Probe light at \SI{313}{\nano\meter} ($s_0 \approx \num{0.03}$) optically pumps excited state atoms to the $\mathrm{{}^3D_J}$ levels, which depletes the \gls{MOT} atom number via decay to the $\mathrm{{}^3P_{0,2}}$ and $\mathrm{{}^1P_1}$ states that are dark to the cooling and imaging light.

\begin{figure}
 \includegraphics[width=85mm]{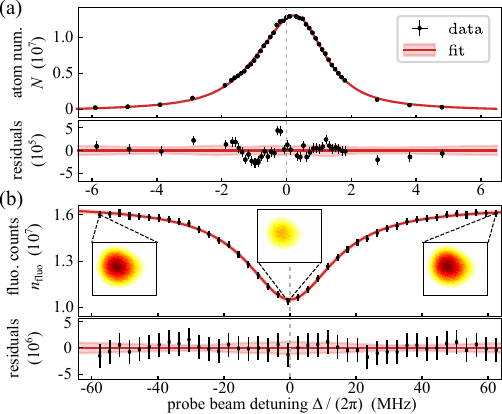}
 \caption{
   \label{fig:2}
   Exemplary spectroscopy data on the (a) $\term{1}{S}{0}{\to}\term{3}{P}{1}$ (\isotope[200]{Hg}) and (b) $\term{3}{P}{1}{\to}\term{3}{D}{1}$ (\isotope[202]{Hg}) transitions in \gls{Hg}.
   Atom numbers $N$ and fluorescence counts $n_\text{fluo}$ extracted from \gls{CCD} images (see inset); the error bars include all systematic shifts and uncertainties.
   The $\term{1}{S}{0}{\to}\term{3}{P}{1}$ fit model takes into account the asymmetry induced by the recoil shift of scattered probe beam photons.
   Minor residual deviations arise from a simplified steady-state optical Bloch solution neglecting Doppler broadening, which does not affect the line center.
   Dashed vertical lines mark the extracted atomic resonances.
  }
\end{figure}

In total, we record approximately \num{100} spectroscopy traces per transition, with about \num{40} frequency steps each, over the course of a few weeks.
Exemplary spectroscopy curves are shown in Fig.~\ref{fig:2}.
The model used for fitting the $\term{1}{S}{0}{\to}\term{3}{P}{1}$ takes into account the recoil shift induced slight asymmetric lineshape.
For the $\term{3}{P}{1}{\to}\term{3}{D}{J}$ data, we fit a simple exponential model, which takes into account a small linear temporal drift of the atom number during the long measurement sequence.
The lineshape models and systematic shifts including light and (residual) Zeeman shifts are discussed in appendices~\hyperlink{appendix:A}{A} and \hyperlink{appendix:B}{B}.

\textit{Results---}Taking into account all systematic shifts and correlations, we obtain the \gls{IS} as $\delta\nu^{A-A'} = \delta_\text{cav}^{\star A} - \delta_\text{cav}^{\star A'}\hspace{-0.17em}$, given in Tab.~\ref{tab:isotope_shifts}.
We resolve the \gls{IS} at \SIrange{2}{11}{\percent} of the natural linewidth, limited by systematic uncertainties. 
Uncertainties are lowest for highly abundant \isotope[198]{Hg} and \isotope[202]{Hg} and increase for less abundant pairs. 
Correlations show weak positive trends due to common systematic corrections.
The \gls{IS} between neighboring pairs ($A'\,{=}\,A\,{+}\,2$) averages to \SI{4.87289(2)}{\giga\hertz} for the intercombination and \SI{-361.7(2)}{\mega\hertz} (\SI{-343.0(2)}{\mega\hertz}) for the two $\term{3}{P}{1}{\to}\term{3}{D}{1,\, 2}$ lines.
It increases approximately linearly with mass number, driven by the \gls{FS}, while the \gls{MS} remains approximately constant \cite{Zhang2019}.
The exception being $\delta\nu^{204-202}$ near the neutron shell closure, where the nucleus is sensitive to onsets of \glsxtrfullpl{ND} that induce the charge radii kink via \gls{FS} contributions $\propto \delta\langle r^2\rangle$ \cite{DayGoodacre2017}.

For the $\term{1}{S}{0}{\to}\term{3}{P}{1}$ line, we improve measurement sensitivity by one to two orders of magnitude compared to Refs.~\cite{Witkowski2019, Schweitzer1963}.
Latest comb-referenced Doppler-free saturation spectroscopy measurements in Refs.~\cite{Witkowski2019, Gravina2025} agree with our results to within $\sim 2\sigma$. 
To our knowledge, no prior \gls{ISS} data or theoretical predictions exist for comparison with the $\term{3}{P}{1}{\to}\term{3}{D}{J}$ spectroscopy data.

\begin{figure}[b]
 \includegraphics[width=85mm]{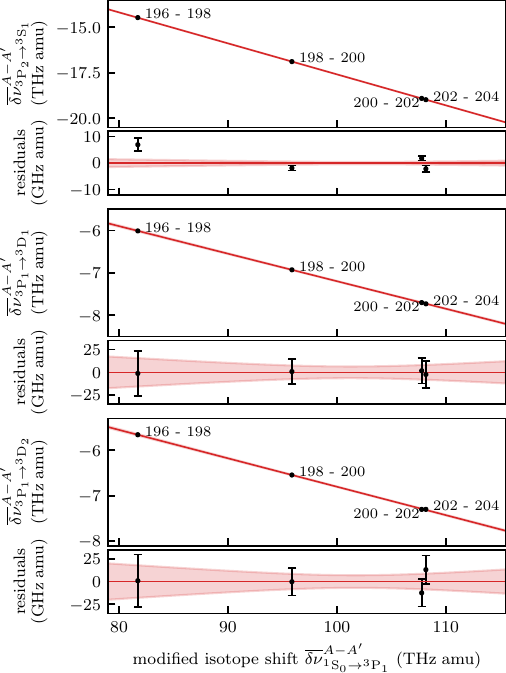}
 \caption{
   \label{fig:3}
   Two-dimensional \glsxtrlong{KP} analysis of the mass-normalized \glsxtrlongpl{IS} $\overline{\delta\nu}_i{}^{A-A'} = \delta\nu_i^{A-A'} / \mu^{A-A'}\hspace{-0.17em}$ with isotope pairs $A'=A+2$, revealing a $4.9\sigma$ nonlinearity in the $\term{1}{S}{0}{\to}\term{3}{P}{1} \rightleftharpoons \term{3}{P}{2}{\to}\term{3}{S}{1}$ comparison.
   Residuals are orthogonal distances to the linear fit.
  }
\end{figure}

\begin{table*}
  \centering
  \setlength{\tabcolsep}{1pt}

  \begin{tabular}{@{}cS[table-format=3.2(2)]S[table-format=4.7(1)]S[table-format=+9.0(2), retain-zero-uncertainty=true, group-separator = {\,}, retain-explicit-plus]S[table-format=+5.2(2), retain-zero-uncertainty=true, group-separator = {\,}, retain-explicit-plus]S[table-format=+5.2(2), retain-zero-uncertainty=true, group-separator = {\,}, retain-explicit-plus]S[table-format=+2.3, retain-zero-uncertainty=true, group-separator = {\,}, retain-explicit-plus]S[table-format=+4.2(2), retain-zero-uncertainty=true, group-separator = {\,}, retain-explicit-plus]S[table-format=+4.2(2), retain-zero-uncertainty=true, group-separator = {\,}, retain-explicit-plus]@{}}
  \toprule
  \multirow{3}*{{$A$}} &
  {\multirow{3}*{{$p_\text{nat}$ (\si{\percent})}}} &
  {\multirow{3}*{{$m$ (u)}}} &
  \multicolumn{4}{c}{$\delta\nu^{A-198}_{\term{1}{S}{0}{\to}\term{3}{P}{1}}$} & 
  {$\delta\nu^{A-198}_{\term{3}{P}{1}{\to}\term{3}{D}{1}}$} &
  {$\delta\nu^{A-198}_{\term{3}{P}{1}{\to}\term{3}{D}{2}}$} \\
  \cmidrule(lr){4-7}
  \cmidrule(lr){8-8}
  \cmidrule(lr){9-9}
  {} &
  {} &
  {} &
  {This work} &
  {Ref.~\cite{Witkowski2019}} &
  {Ref.~\cite{Schweitzer1963}} &
  {Ref.~\cite{Schelfhout2022}} &
  {This work} &
  {This work} \\
  {} &
  {} &
  {} &
  {(\si{\kilo\hertz})} &
  {(\si{\mega\hertz})} &
  {(\si{\mega\hertz})} &
  {(\si{\giga\hertz})} &
  {(\si{\mega\hertz})} &
  {(\si{\mega\hertz})} \\
  \midrule
  {196} &  0.15(1)  & 197.587246(4)    &
       +4179946(30) &
              {---} &          {---}   &
             +4.406 &
        -307.49(98) &      -289.4(1.2) \\
  {198} &  9.97(20) & 199.6045762(7)   &
              0(0)  &
              0(0)  &           0(0)   &
              0     &
              0(0)  &         0(0)     \\
  {200} & 23.10(19) & 201.6219061(9)   &
       -4805516(36) &
       -4806.28(33) &      -4805(4)    &
             –5.054 &
        +347.24(58) &      +328.02(62) \\
  {202} & 29.86(26) & 203.6392358(12)  &
      -10101224(22) &
      -10101.74(28) &      -10102(4)   &
            –10.579 &
        +725.54(58) &      +686.72(61) \\
  {204} &  6.87(15) & 205.6565658(8)   &
      -15311605(36) &
      -15312.02(30) &      -15312(13)  &
            –16.060 &
       +1097.82(58) &     +1038.31(63) \\
  \midrule\midrule
  {} & {} & {\multirow{1}*{{$F$ (\si{\giga\hertz\per\femto\meter\squared})}}} &
  \multicolumn{1}{S[table-format=+2.2(2), retain-explicit-plus]}{-52.22(30){${}^\ast$}} &
  \multicolumn{1}{S[table-format=+2.1(3), retain-explicit-plus]}{-52.7(5.3)} &
  {} &
  \multicolumn{1}{S[table-format=+2.2(4), retain-explicit-plus]}{–58.07(1.00)} &
  \multicolumn{1}{S[table-format=+1.3(2), retain-explicit-plus]}{+3.370(43){${}^\ast$}} &
  \multicolumn{1}{S[table-format=+1.3(2), retain-explicit-plus]}{+3.243(48){${}^\ast$}} \\
  {} & {} & {} & 
  \multicolumn{1}{S[table-format=+2.2(2), retain-explicit-plus]}{} &
  {} & {} & {} & 
  \multicolumn{1}{S[table-format=+1.3(2), retain-explicit-plus]}{+3.42(35){${}^\dagger$}} & 
  \multicolumn{1}{S[table-format=+1.3(2), retain-explicit-plus]}{+3.29(33){${}^\dagger$}} \\
  {} & {} & {\multirow{1}*{{$K$ (\si{\tera\hertz\amu})}}} &
  \multicolumn{1}{S[table-format=+2.2(2), retain-explicit-plus]}{-2.43(55){${}^\ast$}} &
  \multicolumn{1}{S[table-format=+2.1(3), retain-explicit-plus]}{4.0(1.2)} &
  {} & 
  \multicolumn{1}{S[table-format=+2.2(4), retain-explicit-plus]}{–1.15(17)} &
  \multicolumn{1}{S[table-format=+1.3(2), retain-explicit-plus]}{-0.585(86){${}^\ast$}} & 
  \multicolumn{1}{S[table-format=+1.3(2), retain-explicit-plus]}{-0.440(95){${}^\ast$}} \\
  {} & {} & {} & 
  \multicolumn{1}{S[table-format=+2.2(2), retain-explicit-plus]}{} &
  {} & {} & {} & 
  \multicolumn{1}{S[table-format=+1.3(2), retain-explicit-plus]}{-0.45(11){${}^\dagger$}} &
  \multicolumn{1}{S[table-format=+1.3(2), retain-explicit-plus]}{-0.31(12){${}^\dagger$}} \\
  \bottomrule
  \end{tabular}
  \caption{
    \Glsxtrlongpl{IS} $\delta\nu^{A-A'}_i$ on the $i = \term{1}{S}{0}{\to}\term{3}{P}{1}$ and $\term{3}{P}{1}{\to}\term{3}{D}{J}$ ($J=1, 2$) transitions as extracted from spectroscopy and in comparison to literature \cite{Witkowski2019, Schweitzer1963, Schelfhout2022} where available.
    Uncertainties represent the combined statistical and systematic uncertainty of the analysis.
    In accordance with literature, only data for $A' = 198$ are shown; full data and correlations are given in Ref.~\cite{ExternalData}.
    We also give the natural abundances $p_\text{nat}$ from Ref.~\cite{Coursey2015} and compute the nuclear masses, $m = M - Z\, m_e + E_b(A, Z)$, with atomic masses $M$ and binding energies $E_b$ taken from Ref.~\cite{Wang2021}.
    The electronic \gls{FS} ($F$) and \gls{MS} ($K$) constants are extracted from \gls{KP} comparisons to  $\boldsymbol{\overline{\delta\langle r^2\rangle}}$ obtained from Ref.~\cite{Angeli2013} ($\ast$) or to $\boldsymbol{\overline{\delta\nu}}_{\term{1}{S}{0}{\to}\term{3}{P}{1}}$ with $F_{\term{1}{S}{0}{\to}\term{3}{P}{1}}$ extracted from theory in Ref.~\cite{Schelfhout2022}~($\dagger$).
  }
  \label{tab:isotope_shifts}
\end{table*}

\textit{King Plot Analysis---}To identify nuclear contributions and other subleading effects, we compare the mass-normalized \glspl{IS} $\overline{\delta\nu}{}_i^{A-A'}\hspace{-0.17em}$ with $\overline{x}{}^{A-A'} \equiv x^{A-A'} / \delta\mu^{A-A'}\hspace{-0.17em}$ of two transitions $i$ and $j$ via a \glsxtrfull{KP}.
For five isotopes and four isotope pairs, the associated \gls{IS} vectors $\boldsymbol{\overline{\delta\nu}}_i$ span a four-dimensional space.
Eliminating $\boldsymbol{\delta\langle r^2\rangle}$ in Eq.~\eqref{eq:isotope_shift}, the \gls{KP} then yields a linear relation between \glspl{IS} on lines $i$ and $j$ to leading order,
\begin{equation}
  \underbracket[0.5pt]{\boldsymbol{\overline{\delta\nu}}_i = F_{ij}\, \boldsymbol{\overline{\delta\nu}}_j + K_{ij}\, \boldsymbol{\mathbb{1}}}_\text{linear KP} + \sum_{\kappa} G_{ij}^{(\kappa)}\, \boldsymbol{\overline{\delta\eta}}_{(\kappa)}\,,
  \label{eq:king_plot}
\end{equation}
where $\boldsymbol{\mathbb{1}} = (1, 1, 1, 1)^\intercal$ denotes the vector of all ones, $F_{ij} = F_i/F_j$ and $X_{ij} = X_i - X_j\, F_{ij}$ for $X \in [K, G]$.
Deviations from linearity reveal higher-order nuclear effects or \gls{BSM} contributions.
The separation of nuclear and electronic factors allow for comparison of vastly different electronic transitions, including comparisons to muonic atoms \cite{King1984} and (highly-charged) ions \cite{Rehbehn2023, Wilzewski2025}.

Given our and one additional \gls{ISS} dataset for the $\term{3}{P}{2}{\to}\term{3}{S}{1}$ line from Ref.~\cite{Rayman1989}, six distinct two-dimensional \glspl{KP} can be constructed; three of which are shown in Fig.~\ref{fig:3}.
We apply orthogonal distance regression on the mass-normalized \glspl{IS} $\overline{\delta\nu}{}^{A-A'}\hspace{-0.17em}$ of neighboring isotope pairs ($A' = A + 2$) to check for \glsxtrshort{KP}-linearity.
A pronounced nonlinearity is observed for the $\term{1}{S}{0}{\to}\term{3}{P}{1}$ to $\term{3}{P}{2}{\to}\term{3}{S}{1}$ \gls{KP}, corresponding to a $4.9\,\sigma$ rejection of the linear model ($\chi^2_\text{red}=14.7$). 
In contrast, the $\term{3}{P}{1}{\to}\term{3}{D}{1,\, 2}$ to $\term{1}{S}{0}{\to}\term{3}{P}{1}$ ($\chi^2_\text{red} = 0.04$ and $1.00$) and to $\term{3}{P}{2}{\to}\term{3}{S}{1}$ ($\chi^2_\text{red} = 0.04$ and $1.12$) \glspl{KP} are all consistent with linearity \footnote{The low $\chi^2_\text{red}$ values indicate overfitting and therefore no statistical tension to the linear model within the uncertainties.}.
Any potential higher-order nuclear or \gls{BSM} effects align to the vectors of nuclear charge radius, $\boldsymbol{\delta{\langle r^2 \rangle}}$, or inverse nuclear mass difference, $\boldsymbol{\delta\mu}$, within the measurement uncertainty. 
Minor common residual deviations visible at $(A,A'){\,=\,}(204,202)$ could again hint at an onset of \glspl{ND} towards the neutron rich isotopes. 
\gls{KP} linearity is also indicated for the $\term{3}{P}{1}{\to}\term{3}{D}{1}$ to $\term{3}{P}{1}{\to}\term{3}{D}{2}$ \gls{KP} ($\chi^2_\text{red} = 0.74$), which does not involve \gls{FS}-sensitive s-states.

We extract the electronic \gls{FS} ($F_i$) and \gls{MS} ($K_i$) factors for the $\term{3}{P}{1}{\to}\term{3}{D}{J}$ lines from the slopes ($F_{ij}$), intercepts ($K_{ij}$) and $F_{\term{1}{S}{0}{\to}\term{3}{P}{1}}$ given from electron structure calculations in Ref.~\cite{Schelfhout2022}.
Both lines show similar values, likely due to dominance of inner-shell contributions.
We additionally extract $F_i$, $K_i$ for all lines from \gls{KP} comparisons of the \glspl{IS} to the mass-normalized nuclear charge radii differences, $\boldsymbol{\overline{\delta\nu}_i} \approx F_i\, \boldsymbol{\overline{\delta \langle r^2 \rangle}} + K_i\,\boldsymbol{\mathbb{1}}$, a direct result of Eq.~\eqref{eq:isotope_shift}.
The charge radii are taken from optical, x-ray and muonic \glspl{IS}, and electron scattering data in Ref.~\cite{Angeli2013}.
The resulting $\term{3}{P}{1}{\to}\term{3}{D}{J}$ \glspl{KP} show good agreement with linearity.
In contrast, the $\term{1}{S}{0}{\to}\term{3}{P}{1}$ and $\mathrm{\term{3}{P}{2}{\to}\term{3}{S}{1}}$ \glspl{KP} show small deviations ($\chi^2_\text{red} = 2.54$, $3.14$), suggesting higher-order, s-state enhanced nuclear effects.
The extracted $F_i$ and $K_i$ (see Tab.~\ref{tab:isotope_shifts}) agree reasonably with theory~\cite{Witkowski2019, Schelfhout2022}.

\begin{figure}[t]
 \includegraphics[width=85mm]{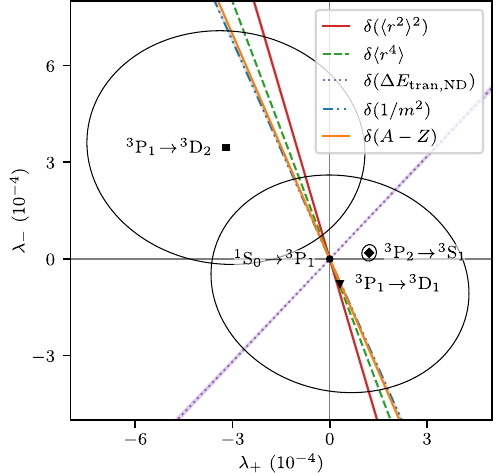}
 \caption{
   \label{fig:4}
   \Glsxtrlong{GKP} nonlinear decomposition analysis of the mass-normalized \glsxtrlongpl{IS} vectors $\boldsymbol{\overline{\delta\nu}_i}$ projected via $\left(F_{ij},\, K_{ij},\, \lambda_+,\, \lambda_- \right) =    \left(\boldsymbol{\overline{\delta\nu}}_j \;\; \boldsymbol{\mathbb{1}} \;\;     \boldsymbol{\Lambda}_+ \;\; \boldsymbol{\Lambda}_- \right)^{-1} \cdot\, \boldsymbol{\overline{\delta\nu}}_j$ via \glspl{IS} of reference line $j = \term{1}{S}{0}{\to}\term{3}{P}{1}$ and vectors~$\boldsymbol{\Lambda}_\pm$ given by Eq.~\eqref{eq:projection_vectors}.
   Higher order nuclear and \gls{BSM} physics contribution vectors, $\boldsymbol{\delta\eta}_{(\kappa)}$, show characteristic nonlinearity patterns that form lines through the origin with slopes $\lambda_- / \lambda_+ =$ 
   \num{-3.418(28)} for \gls{QFS} $\propto \delta\langle r^2 \rangle^2$ \cite{Angeli2013},
   \num{-3(69)} for \gls{ND} $\propto {\delta\langle r^4 \rangle}$ \cite{Shang2024}, 
   \num[retain-explicit-plus]{+1.061(15)} for \gls{ND} $\propto {\delta (\Delta E_\mathrm{tran, ND})}$ from highly-charged ion calculations \cite{Sun2024}, 
   \num{-2.2542(5)} for 2\textsuperscript{nd}-order \gls{MS} $\propto {\delta(1/m^2)}$ \cite{Shang2024} and \num{-2.3227(7)} for \gls{BSM}-coupling to a new boson $\propto (A - Z)$.
   The $1\sigma$ uncertainty bands (shaded) are partially within the linewidth and omitted for \gls{ND} $\propto {\delta\langle r^4 \rangle}$ to improve visibility.
  }
\end{figure}

\textit{Generalized King Plot---}To distinguish between higher-order nuclear and \gls{BSM} contributions, we analyze the nonlinearity in a \gls{GKP} \cite{Hur2022, Berengut2025}.
Following Eq.~\eqref{eq:king_plot}, two additional basis vectors $\boldsymbol{\Lambda}_{\pm}$, independent of the two linear \gls{KP} directions ($\boldsymbol{\Lambda}_\pm \notin \operatorname{span}(\boldsymbol{\overline{\delta\nu}}_j,\boldsymbol{\mathbb{1}})$), are sufficient to span the vector space.
Any \gls{IS} vector can then be expressed as the projection
\begin{equation}
    \boldsymbol{\overline{\delta\nu}}_i = F_{ij}\, \boldsymbol{\overline{\delta\nu}}_j + K_{ij}\, \boldsymbol{\mathbb{1}} + \lambda_+\, \boldsymbol{\Lambda}_+ + \lambda_-\, \boldsymbol{\Lambda}_- \,.
    \label{sec:generalized_king_plot_projection}
\end{equation}
Following the definition in Ref.~\cite{Wilzewski2025}, we choose 
\begin{equation}
  \begin{split}
    \boldsymbol{\Lambda}_+ &= ({\displaystyle 
    \overline{\delta\nu}{}^{c}_j{-}\overline{\delta\nu}{}^{b}_j,\,
    \overline{\delta\nu}{}^{a}_j{-}\overline{\delta\nu}{}^{d}_j,\,
    \overline{\delta\nu}{}^{d}_j{-}\overline{\delta\nu}{}^{a}_j,\,
    \overline{\delta\nu}{}^{b}_j{-}\overline{\delta\nu}{}^{c}_j
    })^\intercal\,, \\
    \boldsymbol{\Lambda}_- &= ({\displaystyle 
    \overline{\delta\nu}{}^{d}_j{-}\overline{\delta\nu}{}^{b}_j,\,
    \overline{\delta\nu}{}^{a}_j{-}\overline{\delta\nu}{}^{c}_j,\,
    \overline{\delta\nu}{}^{b}_j{-}\overline{\delta\nu}{}^{d}_j,\,
    \overline{\delta\nu}{}^{c}_j{-}\overline{\delta\nu}{}^{a}_j
    })^\intercal\,,
  \end{split}
  \label{eq:projection_vectors}
\end{equation}
with isotope pairs $a = (196,\, 198)$, $b = (198,\, 200)$, $c = (200,\, 202)$, $d = (202,\, 204)$, and plot the resulting decomposition in Fig.~\ref{fig:4}.
We compare the projections $(\lambda_{+}, \lambda_{-})$ with the characteristic nonlinearity shapes of potential higher-order corrections $\propto \overline{\delta\eta}_{(\kappa)}$, which form lines through the origin.
The distance to the origin is determined by the associated electronic factors $G_{ij}^{(\kappa)}$.

We consider \gls{QFS} $\propto\delta\langle r^2 \rangle^2$, derived from charge radii in Ref.~\cite{Angeli2013}, as a next leading order \gls{FS} contribution.
While compared to \gls{Yb}, nuclear quadrupole moment deformations in \gls{Hg} are estimated to be small ($\beta_2 \approx \num{0.094(1)}$~\cite{Pritychenko2016}) we also include energy corrections from a quadrupole-deformed Fermi model based on electron structure calculations of highly-charged ions, $\propto {\delta (\Delta E_\mathrm{tran, ND})}$, given in Ref.~\cite{Sun2024}.
A rough estimate \footnote{We assume a charge radius uncertainty of \SI{0.02}{\femto\meter} from Fig.~2 in Ref.~\cite{Shang2024} and propagate the uncertainty to $\delta\langle r^4\rangle$.} of \gls{ND}-induced contributions $\propto \delta \langle r^4 \rangle$ is taken from Ref.~\cite{Shang2024}, where charge radii moments of Hg are generated by a deep neural network trained on relativistic mean-field models and experimental data.
Second-order \gls{MS} $\propto {\delta(1/m^2) = 1/({m^A})^2 - 1/({m^{A'}})^2}$ is expected to be suppressed, but is also hard to distinguish from \gls{BSM} interactions $\propto (A - Z)$ in the \gls{GKP}.
At the current measurement resolution, \gls{GKP} analyses of the $\term{3}{P}{1}{\to}\term{3}{D}{J}$ data do not restrict any nonlinearity source.
The $\term{3}{P}{2}{\to}\term{3}{S}{1}$ to $\term{1}{S}{0}{\to}\term{3}{P}{1}$ point however reproduces its $4.9\sigma$-nonlinearity, that is not easily explained by theory predictions.
While \gls{QFS}, 2\textsuperscript{nd}-order \gls{MS} and \gls{BSM} coupling can be rejected as a single nonlinearity source, multiple sources including \gls{ND} or \gls{NP} could still explain the observed nonlinearity.

\textit{Summary and Outlook---}This work reports first observation of \gls{KP} nonlinearity in neutral \gls{Hg} at the \SI{20}{\kilo\hertz} level, revealing sensitivity to higher‑order nuclear effects relevant for tests of fundamental interactions.
Although current precision and theory predictions limit the \gls{GKP} interpretation, advancing electronic structure calculations alongside refined \gls{ND} predictions, as demonstrated for \gls{Yb}, will enable nuclear $G_i^{(\kappa)}$ and new physics $X_i$ coupling estimates, turning \gls{Hg} \gls{ISS} into a measurement of nuclear structure and a firm bound on \gls{BSM} physics.

Future gain in sensitivity will come from improved spectroscopy resolution on existing transitions or adding at least one (dipole-allowed) line with comparable uncertainty.
\gls{BSM} searches will strongly benefit from accessing ultra-narrow clock transitions such as the $\term{1}{S}{0}{\to}\term{3}{P}{0,\,2}$ lines at \SI{265.6}{\nano\meter} and \SI{227}{\nano\meter} \cite{Tyumenev2016, Porsev2017, Alden2014, Taichenachev2006, Santra2005}.
These dipole-forbidden transitions, weakly enabled via hyperfine or field-induced mixing, offer \si{\hertz}-level linewidths and could be driven either directly or via multi-photon schemes \cite{Alden2014} with available laser technology.  
This also opens prospects to a closed \gls{IS} cycle ($\term{1}{S}{0}{\to}\term{3}{P}{1}$, $\term{3}{P}{1}{\to}\term{3}{S}{1}$, $\term{3}{S}{1}{\to}\term{3}{P}{2}$, and $\term{3}{P}{2}{\to}\term{1}{S}{0}$) that would provide stringent internal tests of \glsxtrshort{SM} and \gls{BSM} interpretations. 

\textit{Note added---}While finalizing this manuscript, we became aware of related work by Gravina \textit{et al.}~\cite{Gravina2025}, which reports similar measurements on $\term{1}{S}{0}{\to}\term{3}{P}{1}$ using a different approach. Our study was conducted independently. 

\textit{Acknowledgments---}We thank {Q.~Lavigne} for contributions to the early stages of the experimental setup and spectroscopy measurements.
We thank B.~Ohayon, J.~Berengut, S.~Heider, and A.~Reu{\ss} for fruitful discussions.
Financial support from the DFG through SFB/TR185 ``OSCAR'' (Project No.~277625399), as well as from the ERC ``quMercury'' (Project No.~757386) and ``UVQuanT'' (Project No.~101080164) is gratefully acknowledged. 

\textit{Data availability---}The data that support the findings of this work are openly available \cite{ExternalData}.
Additional data are available upon reasonable request from the authors.

\appendix
\vspace{1em}

\hypertarget{appendix:A}{\textit{Appendix~A: Systematics---}}The systematic shifts and their corrections for the determination of the atomic resonances with respect to the cavity lines, $\delta_\text{cav}^{\star A}$, are mostly common mode between isotopes and cancel in the computation of the \glspl{IS}.
The dominant sources of systematic shift are probe recoil induced Doppler shifts, AC Stark light shifts, Zeeman shifts, and the slow drift of the reference cavity.
Uncertainties quoted in Tab.~\ref{tab:systematics_combined} are conservative upper bounds, due to correlations or limited ability to isolate individual effects.

\begin{table}[htp]
  \centering
  \begin{tabular}{
    @{}
    c
    l
    l
    S[table-format=4(3),retain-zero-uncertainty=true]
    @{}
    }
    \toprule
     & Systematic & Correction & {Shift (\si{\kilo\hertz})} \\
    \midrule
    \parbox[t]{6mm}{\multirow{8}{*}{\rotatebox[origin=c]{90}{$\term{1}{S}{0}{\to}\term{3}{P}{1}$}}}
    & Probe recoil & $a_I^A\, I_\text{probe}$ & 33(15) \\
    & Zeeman shift & numerical model & -5(15) \\
    & Reference drift & $4\, a_\text{cav}\, t$ ($t\approx\SI{22}{days}$) & 1517(34) \\
    & Probe frequency & $\pm\Delta f_\text{probe}$ & 0(22) \\
    & Doppler shift & $\pm\Delta f_\text{Doppler}$ & 0(20) \\
    & Light shift & $\pm\Delta f_\text{light}$ & 0(0) \\
    & $N$ fluctuations & $\pm\Delta N$ & 0(15) \\
    & $N$ determination & $\pm\Delta N_{od}$ & 0(5) \\
    \midrule
    \parbox[t]{6mm}{\multirow{7}{*}{\rotatebox[origin=c]{90}{$\term{3}{P}{1}{\to}\term{3}{D}{J}$}}}
     & Light shift & $\delta\nu_\text{light}(\Delta_\text{mot},\, P_\text{mot})$ & 3900(170) \\
     & & & -560(810){${}^\dagger$} \\
     & Zeeman shift & $a_B\, \partial_z B + b_B$ & 0(220) \\
     & Probe recoil & $\delta\nu_\text{probe}$ & 0(150) \\
     & Reference drift & $5\, a_\text{cav}\, t$ ($t\approx\SI{9}{\hour}$) & 26(1) \\
     & Probe frequency & $\pm\Delta f_\text{probe}$ & 0(15) \\
     & $N$ drift & $\pm\Delta N$ & 0(70)\\
    \bottomrule
  \end{tabular}
  \caption{
    Upper limits for the systematic shifts and uncertainties in determination of the $\term{1}{S}{0}{\to}\term{3}{P}{1}$ and $\term{3}{P}{1}{\to}\term{3}{D}{J}$ ($J=1,2$) resonance locations, $\delta_\text{cav}^{\star A}$.  
    Uncertainties represent $1\sigma$ combined standard uncertainties.
    Entries marked with $\dagger$ apply only to isotope \isotope[196]{Hg}.
  }
  \label{tab:systematics_combined}
\end{table}

We characterize and correct all systematics based on direct calibration measurements, supported by (numerical) models.
Residual recoil and light shifts are corrected with calibration measurements based on probe intensity $I_\text{probe}$ and show small isotope dependence (scaling slightly with nuclear mass).
Off-resonant light shifts are estimated for both probe and \gls{MOT} beams.
For $\term{1}{S}{0}{\to}\term{3}{P}{1}$ (detected in free-fall), stray \gls{MOT} light is negligible due to fast shuttering.
For $\term{3}{P}{1}{\to}\term{3}{D}{J}$ (in-\gls{MOT} spectroscopy), \gls{MOT}-induced shifts and fluctuations are measured as a function of beam power and detuning.

Zeeman shifts arise from (residual) magnetic gradient fields; for $\term{3}{P}{1}{\to}\term{3}{D}{J}$ we operate the \gls{MOT} at \SI{\sim 0.1}{\tesla\per\meter}.
Numerical modeling and data analysis confirm shift suppression except for \isotope[196]{Hg} detected at higher MOT gradients.
Cavity drift is corrected by a linear fit of atomic resonance to cavity detuning over time, as discussed above.
Laser lock errors and probe frequency instabilities are conservatively estimated using (in-loop) error signals and error statistics from repeated scans.
Atom number uncertainties and extraction (e.g., optical density fit error) are included as nuisance errors.

Background gas and Hg-Hg collisional shifts are negligible at the low background gas pressure of $< \SI{1e-9}{\milli\bar}$ in the \glsxtrlong{UHV} environment of the \gls{MOT}.
Residual Doppler shifts from atomic motion are minimized by balancing \gls{MOT} beam powers and aligning the probe orthogonal to gravity, limiting systematic shifts to below \SI{20}{\kilo\hertz} and isotope-dependent effects to under \SI{100}{\hertz\per\amu}.  
Other broadening mechanisms, such as Doppler and transit-time effects, do not shift the line center positions and add only minor uncertainty to the atomic resonance extraction.

\hypertarget{appendix:B}{\textit{Appendix~B: Lineshape---}}For fitting of the $\term{1}{S}{0}{\to}\term{3}{P}{1}$ spectroscopy data the lineshape model needs to take into account the asymmetry induced by the recoil shift of scattered probe beam photons.
For this we numerically determine the number of scattered photons $n_\text{scat}$ (4 to 15 photons on resonance, $\Delta = 0$) in dependence of the probe beam detuning~$\Delta$,
\begin{equation}
    n_\text{scat}(\Delta) = \left(n \;\Bigg|\; 
   \!\!\int\limits_{0}^{n-1}\!\!\mathrm{d}i\; \frac{1}{R_\text{scat}(\Delta + i\, \Delta_\text{rec})} \overset{!}{=} t_\text{exp} \right),
  \label{eq:lineshape_1S0_3P1}
\end{equation}
where we assume a photon scattering rate of 
\begin{equation}
    R_\text{scat}(\Delta) = \frac{\Gamma}{2}\, \frac{s_0}{1 + s_0 + (2\, \Delta / \Gamma)^2}\,.
\end{equation}
$\Delta_\text{rec}$ is the single-photon recoil shift.
The spectroscopy fit of $N \propto n_\text{scat}$ versus frequency, extends Eq.~\eqref{eq:lineshape_1S0_3P1} by introducing a variable amplitude and offset.  

For the $\term{3}{P}{1}{\to}\term{3}{D}{J}$ spectroscopy the probe beam depletes the \gls{MOT} atom number.
Assuming probe beam dominated atom loss and neglecting Doppler broadening, the loss rate is given by $R_\text{loss}(\Delta) = \beta_\text{loss}\, R_\text{scat}(\Delta)$ with $\beta_\text{loss}$ the branching ratio to metastable states.
The atom number then decays as $N(t) \propto e^{-R_\text{loss}(\Delta) \, t}$ and the detected fluorescence is given by the integral over the time, $n_\text{fluo}(\Delta) \propto \int_0^{t_\text{exp}}\! N(t)\, \mathrm{d}t$.
To account for slow drift in atom number due to \gls{MOT} cooling power variations, we add a linear background term $\propto a_n \cdot \Delta$, and obtain
\begin{equation}
\begin{split}
  n_\text{fluo}(\Delta)
    = &\left(n_0 + a_n\, \Delta \right)\, \left[1 + \left({2\,\Delta}/{\Gamma}\right)^2 \right]\,\\ &\left[1 - \exp\left( -\frac{a}{1 + (2\,\Delta/\Gamma)^2} \right)\right]\,.
  \label{eq:fit_model_313_with_bg}
\end{split}
\end{equation}
The additional systematic uncertainty of this drift is estimated to be \SI{70}{\kilo\hertz}.
We validate both fit models against pure Lorentzian, Gaussian, and Voigt profiles, which in comparison reduce fit quality and introduce overfitting or parameter correlations.

The fits take into account all systematic shifts, uncertainties, and their correlations.
All confidence interval estimations in this work are performed via the \texttt{kafe2} \texttt{python} toolbox \cite{Gaessler2022} which interfaces the \texttt{Minuit2} \texttt{C++} library maintained by the CERN ROOT team \cite{Brun2019}.

\end{document}